\documentstyle[preprint,aps,prd]{revtex}
\tighten

\begin{document}
\draft
\title{Large nucleation distances from impurities in the cosmological 
quark-hadron transition}
\author{Michael B. Christiansen and Jes Madsen}
\address{Institute of Physics and Astronomy, University of Aarhus,
DK-8000 \AA rhus C, Denmark}
\date{December 4, 1995}
\maketitle

\begin{abstract}
We calculate the mean nucleation distance, $d_{nuc}$, in a first order
cosmological quark-hadron phase transition. For homogeneous nucleation
we find that self-consistent inclusion of curvature energy reduces
$d_{nuc}$ to $\lesssim 2$cm. However, impurities can lead to
heterogeneous nucleation with $d_{nuc}$ of several meters, a value high
enough to change the outcome of Big Bang nucleosynthesis.
\end{abstract}

\pacs{98.80.Cq, 12.38.Mh, 12.39.Ba, 26.35.+c\\ \\ \\
To appear in Physical Review D, May 15, 1996}

\section{INTRODUCTION}
A number of authors \cite{witt,hogan,alc1,kunuk} have
suggested that inhomogeneities
in the baryon number density were produced in the cosmological quark-hadron
phase transition and perhaps even persisted to the time of nucleosynthesis,
which could alter the abundances of light elements. For this to happen,
a few criteria have to be met, though. First of all, only 
a first order quark-hadron transition seems capable of generating
large baryon number fluctuations. Whether the transition is first order
is still an unsettled question (see e.g.\ Ref.\ \cite{iwasaki}). Secondly, 
the mean distance
between high and low baryon density regions, denoted {\it l}, has to be larger
than approximately one meter comoving at $ T=100 $MeV
\cite{kunuk}, if inhomogeneous nucleosynthesis results are to differ 
from the standard homogeneous nucleosynthesis results (otherwise,
neutron and proton diffusion will erase the inhomogeneities prior to
nucleosynthesis).
Other important parameters are the average density contrast between high and
low-density regions, the average volume fraction of high-density
regions, and the baryon to photon ratio.

We assume that the transition is first order and focus in this 
paper on determination of the most 
important of the parameters, the mean distance $ l $.

A first order quark-hadron transition in the expanding Universe
can in general terms be described as follows.
As the color deconfined quark-gluon plasma cools
below the critical temperature, $ T_c \approx 100$MeV, it
becomes energetically
favorable to form color confined hadrons (primarily pions and a 
tiny amount of neutrons and protons due to the conserved net baryon number).
However, the new phase does not show up 
immediately. As is characteristic for a first order phase transition, some 
supercooling is needed to overcome the energy expense of forming the surface
of bubbles of the new hadron phase. 
When a hadron bubble is nucleated, latent heat is
released, and a spherical shock wave expands into the surrounding supercooled 
quark-gluon plasma. This reheats the plasma to the critical temperature, 
preventing 
further nucleation in a region passed by one or more shock fronts \cite{kuka} 
(for the range of parameters used in the present investigation, 
bubble growth is described by deflagrations, where a 
shock front precedes the actual phase transition front. However, if the shock 
fronts are very weak the quark-gluon plasma may not be reheated enough to 
sufficiently prevent new bubble nucleations in such regions \cite{kuka}. This 
reduces the 
mean distance. We will neglect this possibility, because of the very small 
amounts of supercooling we deal with here). 
The nucleation stops when the whole Universe has 
reheated to $ T_c $. This part of the phase transition passes very fast, in
about $ 0.05\mu $s, during which the cosmic expansion is totally negligible. 
After that, the hadron bubbles grow at the expense of the quark phase
and eventually percolate or coalesce; almost in 
equilibrium at $ T_c $. The transition ends when all quark-gluon matter has been
converted to hadrons, neglecting possible quark nugget production \cite{witt}.

How much the system has to supercool before the first hadron bubbles show up 
determines the distance between nucleation sites. This distance is called the
(comoving) mean nucleation distance, $ d_{nuc} $. It is also roughly
the distance between the shrinking, high-baryon-density quark droplets.  
$ l $ is half that distance, and when we take into account that the Universe 
expanded about $ 50\% $ during the phase transition, we have
\begin{equation}
\label{l}
l \simeq 0.8 \left ( 
\frac{T_c}{100\hbox{MeV}} \right ) d_{nuc},
\end{equation}
where the factor $T_c/100$MeV scales the value of $d_{nuc}$ from a given
critical temperature to $T=100$MeV, where $l$ is defined.

In the bag model
we can approximate $ d_{nuc} $ in homogeneous nucleation theory without 
curvature energy in terms of the bag constant, $B$, by
\begin{equation}
\label{dsurf}
d_{nuc} \approx 13  \left ( \frac {145\hbox{MeV}}{B^{1/4}} \right )^2
\left ( \frac{4B}{L} \right )
\left ( \frac {\sigma }{0.035B^{3/4}} \right )^{3/2} \hbox{cm},
\end{equation}
where $ L $ is the latent heat and $ \sigma $ the surface tension.
Lattice QCD calculations \cite{surft} suggest that 
$ \sigma \lesssim 0.24T_c^3 $, Ref.\ \cite{latent} finds $ \sigma 
\simeq 0.025T_c^3 $ and $L\simeq 0.4B$. With $ T_c \simeq 0.68B^{1/4} $,
this gives $ d_{nuc} \simeq 0.1 $m for $ T_c > 100 $MeV. 

We calculate the 
surface tension self-consistently  within the multiple reflection expansion
framework of the MIT bag model and get $ \sigma \lesssim 0.035B^{3/4} $
and $L\simeq 4B$,
corresponding
to $ d_{nuc} \lesssim 0.1 $m. But we also self-consistently take into account
the important curvature term and find a crude approximate formula for
$ d_{nuc} $, 
\begin{equation}
\label{dcurv}
d_{nuc} \approx 1.8 \left ( \frac{145\hbox{MeV}}{B^{1/4}} \right )^2 
\left ( \frac{4B}{L} \right ) \hbox{cm}.
\end{equation}
Note, that in this approximation the mean nucleation distance depends only on
the bag constant (or $ T_c $). Thus the overall effect of including a 
curvature 
term is roughly a factor of 7 reduction of $ d_{nuc} $.
With these estimates of $ d_{nuc} $ in mind it hardly seems possible to get
any significant deviations from standard nucleosynthesis, if homogeneous 
nucleation theory applies. 

However, homogeneous nucleation may not be the prime mechanism
responsible for hadron formation.
It is well-known from every day life that first order transitions are normally 
facilitated by impurities, such as dust or ions and container walls. Think
e.g.\ of a charged particle entering a cloud chamber, or boiling water where 
(half) bubbles are formed at the bottom of the pot. Such nucleations are 
denoted heterogeneous. Thus it is obvious to consider the effect of such
impurities on the quark-hadron phase transition. Candidates are the topological
impurities, like primordial black holes, magnetic monopoles, and cosmic strings
or relic fluctuations from the electroweak transition. 

The idea that impurities could play a role in the cosmological
quark-hadron transition is not new (see e.g.\ \cite{hoso} for a general 
discussion of the role of impurities in cosmological phase transitions), but 
normally it has been
argued\cite{alc1} that the presence of
impurities would just act as extra seeds for nucleation sites together with
the homogeneous nucleation sites and thus reducing $ d_{nuc} $. 

As demonstrated below, this picture is not necessarily correct---there
are circumstances where impurities lead to a dramatic {\it increase\/}
in $d_{nuc}$, maybe even to a value of several meters, resurrecting
inhomogeneous Big Bang nucleosynthesis.
We find that three limits with,
respectively, small, medium and huge amounts of impurities, compared to the
number of homogeneous nucleation sites, have to be considered. Certainly, if 
only very few impurities are present in the early Universe $ d_{nuc} $ will
be determined solely by homogeneous nucleation. In the other extreme,
for a huge number of impurities, they will determine a $ d_{nuc} $
smaller than the one obtained from homogeneous nucleation. So neither of these
cases give large mean nucleation distances. 

Any impurity tends to lower the cost
in energy when creating the surface of the new phase. Therefore bubbles of the
new phase will nucleate easier around an impurity at a smaller amount of 
supercooling. This means that shock spheres from nucleations around impurities
start reheating the Universe earlier. And for some impurity number densities
the whole Universe has already been reheated before the amount of supercooling
has increased to a value where homogeneous nucleation sets in.
In this case $ d_{nuc} $
depends on the impurity number density, $ n $, on the time
when heterogeneous nucleation starts, and on the time when homogeneous
nucleation becomes important. For a large range of parameters,
$d_{nuc}$ can be very big.

We will make this qualitative discussion much
more quantitative in Section III, where we also present a
self-consistent bag model calculation of homogeneous nucleation,
including for the first time the important curvature terms. 
In Sec.\ II we describe the thermodynamics of the
phase transition and the multiple reflection expansion of the MIT bag model,
which we use to calculate the thermodynamical quantities. Section IV contains
a general discussion and a summary of our results.

\section{THERMODYNAMICS AND THE BAG MODEL}
All the necessary thermodynamical information about the two phases is
contained in the thermodynamical
potential $ \Omega (T,\mu ,V,S,C) $. $ \mu $ is the chemical potential, $ V $  
the volume, and finite size corrections enter via the surface area, $ S $, 
and the extrinsic curvature, $ C $.
\begin{equation}
\Omega = -T \ln Z, 
\end{equation}
where $ Z $ is the grand partition function. All other thermodynamical
quantities can be derived from $ \Omega $. For instance, the 
bulk pressure $ P $ is
\begin{equation}
 P = - \left ( \frac{\partial \Omega}{\partial V} \right )_{T,\mu ,S,C},
\end{equation}
and the surface tension $ \sigma $
\begin{equation}
 \sigma = \left ( \frac{\partial \Omega}{\partial S} \right )_{T,\mu ,V,C}.
\end{equation}
Because of the huge photon to baryon ratio, we can neglect all chemical
potentials and simplify the equations for $ \Omega $ when we want to
calculate $d_{nuc}$, since 
$ (\mu /T )\sim 10^{-9} $ in the early Universe (see e.g. \cite{kolb}).
(This is not true if one is interested in following the
further evolution of baryon density contrasts.)

\subsection{The quark phase}
The quark phase consists of massless gluons, massless (or with
masses of 5--10 MeV) $ u $ and $ d $ quarks, and a massive $ s $ quark, 
where we use a broad interval, $ m_s/\hbox{MeV} \in \left [ 50,300 \right ] $. 
In most calculations we have integrated numerically the $s$-quark
distribution function; however for simplicity we neglect the small $s$-quark
contribution to the cosmic density when calculating the cosmological
time-temperature relation from the Friedmann equations.
The gluons contribute with 16 degrees of freedom, while each of the quarks has
12 degrees of freedom (including antiparticles).
In thermal equilibrium with both 
phases during the transition are (statistical weight in parenthesis) 
electrons(4), muons(4), neutrinos(6) and photons(2). The contributions
from these particles cancel when we study pressure differences.

The behavior of quarks is mainly governed by the strong interaction. It can, at
least in principle, be described in the QCD model. But
since QCD is a non-perturbative theory at low energy and therefore {\it very} 
difficult to handle, phenomenological models, like the MIT bag model, 
or lattice QCD calculations are used in practice.
The basic idea of the bag model is that the quarks can be treated as an
(almost) ideal gas confined to a finite region of space, a bag, by what we may
think of as an external pressure, $B$, called the bag constant.
In the multiple reflection expansion framework \cite{balian} a statistical 
approach is taken and finite size effects of the bag are explicitly 
incorporated in the density of states formula Eq.\ (\ref{rho}) 
in terms of geometrical factors.

What value to take for the phenomenological bag constant is at present very 
uncertain; it could even be temperature dependent. A lower limit on the bag 
constant of $ (145{\hbox {MeV}})^4 $ is set by the stability of nuclei, 
especially $ ^{56} $Fe, relative to $ ud $-quark matter. There is no upper 
limit available from physical considerations. Lattice QCD calculations 
typically favor $ B^{1/4} $ above 200MeV. We use here a broad interval for 
the bag constant, $ B^{1/4}/{\hbox {MeV}} \in [145;245] $.
It was shown, in the $ T=0 $ limit, by Farhi and Jaffe \cite{farhi} that a QCD 
coupling constant 
different from zero can largely be absorbed in a reduction of the bag constant.
We assume this is valid at finite temperatures too and 
put $ \alpha_s=0 $. 

The multiple reflection expansion was originally
derived for a quark droplet embedded in a hadron gas. But Mardor and Svetitsky
\cite{mardor} have shown that 
it can also be used to describe a hadron bubble in a quark plasma, 
provided the signs of the volume and curvature terms are changed. The 
formulae below are for a quark droplet. The
thermodynamical potential of the quark phase is then given by
\begin{equation}
 \Omega_{q}= \sum_{i} \Omega_{i} +BV,
\end{equation}
where $ \Omega_{i} $ is the thermodynamical potential for a single fermion 
(quark) or boson (gluon), given as
\begin{equation}
\label{zi}
 \Omega_{i} = -T \ln Z_i.
\end{equation}
Here $ Z_i $ is the corresponding grand partition function
\begin{equation}
 \ln Z_i = \pm \int_0^{\infty} \, dk \, \rho(k) \ln \left 
(1 \pm e^{-\sqrt{k^2+m_i^2}/T} \right )
\end{equation}
(upper sign for fermions, lower for bosons) where the density of states is 
obtained from the multiple reflection expansion
\begin{eqnarray}
\label{rho}
 \rho (k) = & g &\left\{  \frac{k^2V}{2\pi^2} +f_{S}\left 
( \frac{k}{m} \right )k\oint_{S}\,dS   \right.   \nonumber \\ 
  &+&\! \left. f_{C}\left ( \frac{k}{m} \right ) \oint_{S} \left 
( \frac{1}{R_{1}} + \frac{1}{R_{2}} \right )\, dS + ... \right\} , 
\end{eqnarray}
$ g $ is the statistical weight of the particle. The volume term is universal 
and we recognize the finite size effects from the surface area, $S$, in
the second term, and from the extrinsic curvature, $C$, 
in the third term. Higher order 
terms are neglected. Assuming that the quark droplets are spherical, then
$ V=4\pi R^3/3,\, S=4\pi R^2 $ and $ C=8\pi R $.
The coefficient functions $ f_{J}(k/m) $ are for quarks \cite{berger}
\begin{equation}
  f_{S}\left ( \frac{k}{m} \right )=-\frac{1}{8\pi} \left 
[ 1-\frac{2}{\pi}\arctan \frac{k}{m} \right ], \;\;\;\;\;\; 
\lim_{m \rightarrow 0 } f_{S}= 0
\end{equation}
and
\begin{eqnarray}
  f_{C}\left ( \frac{k}{m} \right ) &=&\frac{1}{12\pi^2}\left 
[1- \frac{3}{2} \frac{k}{m}  \left ( \frac{\pi}{2}-\arctan \frac{k}{m} 
\right ) \right ],
 \nonumber \\
 & & \;\;\;\;\; \lim_{m \rightarrow 0 } f_{C}= -\frac{1}{24\pi^2}
\label{curv}
\end{eqnarray}
$ f_{C} $ has not been calculated from the multiple reflection expansion 
for massive quarks, but Eq.\ (\ref{curv}) gives a
perfect fit to shell model calculations for strangelets\cite{mads}.
For the massless gluons we have \cite{mardor}
\begin{equation}
  f_{S}=0, \;\;\;\;\;\;\; f_{C}=-\frac{1}{6\pi^2}.
\end{equation}
Note, that these results for gluons have been confirmed by pure glue QCD 
lattice calculations \cite{gamglu}. 

\subsection{The hadron phase}
We treat the hadron phase as an ideal pion gas, where we calculate the pressure
from $ \Omega_i $ using Eq.\ (\ref{zi}) for bosons and only including the 
volume term in the density of states formula Eq.\ (\ref{rho}). 
This approximation is justified 
because the pions, with masses around the critical temperature 100-200MeV, 
are the lightest hadrons and semi-relativistic while all other hadrons are 
nonrelativistic and therefore contribute negligibly to the pressure for
$ T\lesssim 200$MeV. 
The pion pressure is much smaller than the bag pressure so even the pressure 
from fully relativistic pions will only increase the critical 
temperature one percent.
A perhaps more serious problem is that pions are not ``point-like''
particles and excluded volume effects should be taken into account.
The pions could in principle contribute to the surface term (and curvature 
term as well), we do not know. However, the pions have only 3 degrees of 
freedom, while each of the quarks have 12, so we would expect only minor 
corrections.

\subsection{Hadron bubbles in quark-gluon plasma}
The change in the thermodynamical potential, when a quark matter 
droplet of radius $ R_q $ is formed in a hadron gas, is
\begin{equation}
 \Delta \Omega _{q-droplet}=P_{h}V_{q} + \Omega_{q}(R_{q}).
\end{equation}
Because the photon-lepton gas is in thermal equilibrium with both 
phases, it does not contribute
to $ \Delta \Omega $. Only in pressure ratios is it of importance to include
the photon-lepton contribution.
In the cosmological quark-hadron phase transition hadron bubbles are formed 
from a quark-gluon plasma. Mardor and Svetitsky \cite{mardor} have shown 
that this situation can be described simply by inverting the system,
i.e let $ R_{q} \rightarrow -R_{h} $. This 
inversion affects only the signs of the volume and curvature terms and not 
the absolute value of any of the coefficients. $ \Delta \Omega $ can then be 
written as 
\begin{equation}
  \Delta \Omega _{h-bubble}=-P_{h}V_h+\Omega_{q}(-R_{h}).
\end{equation}
or 
\begin{equation}
 \Delta \Omega = -\frac{4\pi}{3}\Delta P R^3 + 4\pi \sigma R^2 -8\pi \gamma R,
\end{equation}
where we have suppressed the index $ h $ on the hadron bubble radius.  
If we relate the temperature and bag constant as
\begin{equation}
  T\equiv xB^{1/4}, \,\,\,\,\,\,\,\,\,\, T_c\equiv x_c B^{1/4},    
\end{equation}
then from the multiple reflection expansion we have
\begin{equation}
\label{dp}
 \Delta P\equiv P_h-P_q = P_{\pi}+B-P_s-\frac{37}{90}\pi^2 x^4 B ,
\end{equation}
where the pion pressure is (we treat the three pion types alike and 
use a mean mass of $ m_{\pi}=138$MeV)
\begin{equation}
 P_{\pi}=-\frac{3x}{2\pi^2}B\int_{0}^{\infty}du\, u^2 \ln\left 
( 1-e^{-\sqrt{u^2+\bar{c}^2}/x} \right ),\,\,\, \bar{c}=\frac{m_{\pi}}{B^{1/4}}
\end{equation}
and the $ s $ quark pressure
\begin{equation}
  P_s= \frac{6x}{\pi^2}B\int_{0}^{\infty}\,du\, u^2 \ln\left 
( 1+e^{-\sqrt{u^2+\tilde{c}^2}/x} \right ),\,\,\,\,\,\, 
\tilde{c}=\frac{m_s}{B^{1/4}}.
\end{equation}
The critical temperature is found from $\Delta P=0$
(the corresponding $ x_c$ lies between 0.67 and 0.70). The 
surface tension, where only the massive $ s $ quarks contribute, is given by
\begin{eqnarray}
  \sigma =\sigma_s=\frac{3x}{2\pi}B^{3/4} \int_{0}^{\infty}\,du\, 
& u & \left ( 1-\frac{2}{\pi}\arctan \frac{u}{\tilde{c}} \right ) \nonumber \\ 
 & \times & \ln \left ( 1+e^{-\sqrt{u^2+\tilde{c}^2}/x} \right ).
\end{eqnarray}
And finally the curvature coefficient
\begin{equation}
  \gamma = \frac{19}{36}x^2B^{1/2}+\gamma_s,
\end{equation}
where
\begin{eqnarray}
  \gamma_s=-\frac{x}{\pi^2}B^{1/2}\int_{0}^{\infty}\, & du & \left [ 
1-\frac{3}{2}\frac{u}{\tilde{c}} \left (\frac{\pi}{2}-\arctan 
\frac{u}{\tilde{c}} \right ) \right ] \nonumber \\
 & \times & \ln \left ( 1+e^{-\sqrt{u^2+\tilde{c}^2}/x} \right ).
\end{eqnarray}
$ \gamma $ is always larger than zero because of the large positive
contribution from the gluons, even though $ \gamma_s $ can be 
negative depending on the mass of the $ s $ quark. 
So the curvature energy is always negative when making a hadron bubble.

The latent heat per volume, released in a first order transition, is defined
as
\begin{equation}
 L\equiv -T_c\left [ \frac{\partial \Delta P}{\partial T} \right ]_{T_{c}}.
\end{equation}
Eq.\ (\ref{dp}) can easily be differentiated with respect to temperature and 
the latent heat calculated numerically for fixed bag constant and strange
quark mass. In Fig.\ \ref{fig1} we have plotted the latent 
heat as a function of the strange quark mass. And we show the effect
of including higher mesonic states up to 1 GeV 
(the higher mesonic states
have much less influence on the critical temperature, because the bag constant
contributes with about $ 95\% $ of the pressure needed to maintain pressure
equilibrium between the two phases. That is why we can safely disregard heavier
mesons for $ T_c \leq 200 $MeV). It is evident that these states only
have a minor influence on the latent heat. A good approximation is to use
\begin{equation}
 L\simeq 4B
\end{equation}
independent of $ m_s $. We adopt this approximation, but notice that some
lattice QCD calculations suggest \cite{latent} 
that the latent heat is much smaller, 
$ L \simeq 0.4 B $. We will discuss the possible discrepancy later.
 
\section{NUCLEATION THEORY}
In classical nucleation theory the nucleation rate (number of nucleations 
per volume per time) is given by \cite{becker,lang} 
\begin{equation}
\label{nukrate}
  p(T)= C(T) \exp \left ( - \frac {\Delta F_{c}}{T} \right ),
\end{equation}
where $ \Delta F_{c} $ is the minimum work needed to create the smallest 
possible growing bubble, equal to the change in Helmholtz free energy, and 
$ \exp(- \Delta F_{c}/T) $ is the probability of making a growing 
bubble. The pre-exponential factor depends in detail on the model by which the 
bubble growth is described. Csernai and Kapusta \cite{cser} have calculated 
$ C(T) $ in the cosmological case (relativistic particles and almost zero 
baryon number), but without a curvature term, and get  
\begin{equation}
\label{ct}
 C(T) = \frac{16}{3^{5/2}\pi }\frac{\sigma ^{5/2} \lambda ^3}{\xi_q^4 L^2
T^{3/2}} R_c,
\end{equation}
where $ R_c $ is the radius of a critical size bubble and 
$ \xi_q \approx 0.7 $fm is the surface thickness of the bubble. 
$\xi_q$ has to be much smaller than $ R_c $, which it is 
since $ R_c \approx 50 $fm, typically. Finally the shear viscosity $ \lambda $ 
is approximated by
\begin{equation}
  \lambda \simeq \frac{1.12}{\alpha_s^2 \ln \alpha_s^{-1}} T^3 ,
\end{equation}
where $\alpha_s$ is the strong coupling constant.
Fortunately, as we will show, a temperature dependent pre-exponential factor 
like Eq.\ (\ref{ct}) is not important when we want to determine $ d_{nuc} $. 
Thus, the often used dimensional estimate $ C(T)\approx T_c^4 $ is sufficient
for this purpose.

\subsection{Homogeneous Nucleation}
This subsection is for a major part a straight forward 
generalization of the work
done by Fuller, Mathews and Alcock \cite{alc1} to take a varying pre-exponential
factor and, more important, a curvature term into account. Their results
(with the correction mentioned in\cite{alc2}) are
recovered in the limit of constant pre-exponential factor and vanishing
curvature.

The change in Helmholtz free energy is equal to the change in the 
thermodynamical potential in the limit of vanishing chemical potentials. Thus,
\begin{equation}
  \Delta F(R,T)= -\frac {4\pi}{3} \Delta P R^{3} +4\pi \sigma R^{2} 
- 8\pi \gamma R.
\end{equation}
The radius of a critical size bubble is found by putting $ \partial \Delta F/
\partial R $ equal to zero. From this equation we get two critical radii 
\begin{equation}
\label{rc}
  R_{\pm} = \frac {\sigma}{ \Delta P} \left ( 1 \pm \sqrt{1-
\frac {2\Delta P \gamma}{\sigma^{2}}} \right ).
\end{equation}
The smaller radius corresponds to a local minimum in the free energy 
(see Fig.\ \ref{fig2}), the 
larger to a local maximum. As pointed out in \cite{mardor} this local minimum
has some serious consequences, if it is real and not just a shortcoming of 
the multiple reflection expansion of the bag model, 
since it means that small hadron bubbles show
up even above the critical temperature. It has been argued that these hadron
bubbles are unstable \cite{lana}, but at present it is unclear how to handle
them. We will for now neglect this complication, keeping it in mind for 
later discussion.

A linear expansion of $ \Delta P $ about the critical temperature gives 
\begin{equation}
 \Delta P = L \eta,
\end{equation}
where $ L$ is the latent heat per volume,
and the supercooling parameter for $ T \leq T_c $ is
\begin{equation}
 \eta \equiv \frac{T_{c}-T}{T_{c}}.
\end{equation}
This expansion is valid for the small supercooling $  \eta  \ll 1 $ we deal
with in the cosmological quark-hadron transition. We then write $\Delta F_{c}= F(R_{+},T)-F(0,T) $ as
\begin{equation}
\label{dfc}
  \Delta F_{c}(\eta)=\frac {8\pi}{3} \frac {\sigma^{3}}{L^{2}\eta^{2}} 
\left ( 1-\frac {3L\gamma}{\sigma^{2}}\eta +
(1-\frac {2L\gamma}{\sigma^{2}}\eta)^{3/2} \right ).
\end{equation}
Certainly $ \Delta F_{c} $ can not be allowed to become negative if we insist
on ignoring the minimum. This gives an upper limit on the supercooling 
parameter,
\begin{equation}
 \eta \leq \frac {3}{8} \frac{\sigma^{2}}{L \gamma}.
\end{equation}

Combining equations (\ref{nukrate}) and (\ref{dfc}) gives the nucleation rate as
a function of the small supercooling parameter, valid around the critical 
temperature:
\begin{equation}
\label{peta}
  p(\eta) = C(\eta)\exp\left [-\frac{8\pi}{3} \frac{\sigma^{3}}{L^{2}T_{\rm c}
\eta^{2}} \left ( 1-3b\eta+ \left(1-2b\eta \right )^{3/2} \right )\right ]
\end{equation}
where
\begin{equation}
 b\equiv \frac{L\gamma}{\sigma^2}.
\end{equation}
We see that $ p(\eta) $ equals zero as we pass the critical temperature from 
above and 
increases very rapidly with $ \eta $. As a result nearly all nucleation takes 
place at the lowest temperature, $ T_{f} $, achieved during the supercooling 
phase. This means that we may generally approximate 
the pre-exponential factor with
its value at $ T_{f} $ provided it is not {\it too\/} temperature dependent.

It turns out to be fruitful to make a linear expansion of $ \ln p (\eta) $ 
about
the time $ t_f $ corresponding to the temperature $ T_f $
\begin{equation}
  \ln p(t) \simeq \ln p(\eta_{f}) + \left [ \frac {d \ln p }{d \eta } 
\right ]_{\eta_{f}} \left [ \frac{d\,\eta}{d\,T} \right ]_{T_f} 
\left [ \frac{d\, T}{d\, t} \right ]_{t_f} (t-t_f),
\end{equation}
or
\begin{equation}
\label{paft}
 p(t)\simeq p(\eta_f) \exp [-\alpha(t-t_f) ].
\end{equation}
This is a good approximation because of the steepness of the nucleation rate
(cf.\ Eq.\ (\ref{peta})).
We have implicitly assumed that the surface tension and curvature coefficient 
are independent of temperature. This is not true in the bag model. But there 
is nothing special at all about the critical temperature from the point of 
view of $ \sigma $ and $ \gamma $. And since the amount of supercooling is 
very small we treat them as constants and use their values at $ T_{c} $. 

The temperature as a function of time is found by solving the Friedmann 
equations for a flat Universe consisting of a quark-gluon plasma, photons and
leptons. It is, in the
first order approximation we use, independent of the bag constant,
\begin{equation}
\label{fried}
  T \simeq \left (\frac{45}{16\pi^{2}g_qG} \right )^{1/4}t^{-1/2},
\end{equation}
where $G$ is the gravitational constant, and the statistical weight
$g_q=51.25$ (for simplicity we here neglect the small contribution from
the semirelativistic $s$-quarks).
$ \alpha $ is then
\begin{eqnarray}
\label{alfa}
  \alpha & = & (41\pi G)^{1/2}   \left( \frac{\pi}{3}T_{c}^{2}   
\frac{C'(\eta_f)}{C(\eta)}  \right.  \\ \nonumber & + & \left. 
\frac{8\pi^{2}}{9} \frac{\sigma^{3}T_{c}}{L^{2}\eta_{f}^{3}} 
\left [ 2-3b\eta_{f}+(2-b\eta_{f})(1-2b\eta_{f})^{1/2} \right ] \right),
\end{eqnarray}
where $C'\equiv dC/d\eta$.

The mean nucleation distance is found from the total number density of 
nucleation sites
\begin{equation}
\label{nnuc}
 n_{nuc} \equiv d_{nuc}^{-3} = \int_{t_c}^{\infty} f(t)p(t) \, dt,
\end{equation}
where $ f(t) $ is the unaffected fraction of the Universe.           

After a hadron bubble is nucleated a spherical shock front expands
into the quark phase at a speed slightly above the sound speed, $ v_s \simeq
3^{-1/2} $, and reheats it to the critical temperature, 
which prevents any further nucleation in a region crossed by 
one or more shock fronts\cite{kuka}. We use the formula
given by Guth and Tye \cite{guth} for the fraction of space not yet passed by
one or more shock fronts (a slightly different approach
can be found in \cite{cser}), 
\begin{equation}
  f(t) = \exp\left [-\int_{t_c}^{t} dt'f(t')p(T(t'))V(t',t)\right ]
\end{equation}
where $t_{\rm c}$ is the time corresponding to temperature $T_{\rm c}$ 
and $V(t',t)$ in this context is the reheated volume at time $t$ caused 
by a bubble nucleated at time $t'$. $V(t',t)$ can be written as 
\begin{equation}
  V(t',t) \approx \frac{4\pi}{3}v_{\rm s}^3(t-t')^3.
\end{equation}
For the short timescales involved here, we can safely neglect the cosmic
expansion. This recursive formula for $ f(t) $ can be simplified to a step 
function by a linear
expansion, giving
\begin{equation}
\label{faft}
  f(t) \simeq \left\{ \begin{array}{ll}
  1-\frac{4\pi}{3} v_{s}^3 \int_{t{c}}^{t}\,dt'\,p(T(t')) (t-t')^3 & t< t_{f} \\
  0 & t \geq t_{f} 
                    \end{array} \right. .
\end{equation}
The time $ t_{f} $ corresponds to the maximum supercooling $ \eta_{f} $, where 
we demand the whole Universe to be reheated. The approximation succeeds because
of the very rapid rise of the nucleation rate. If we introduce the 
approximation for the nucleation rate, Eq.\ (\ref{paft}), the integral in 
Eq.\ (\ref{faft}) can be analytically evaluated and 
\begin{equation}
\label{fafta}
  f(t) \simeq \left\{ \begin{array}{ll}
  1- \frac {8 \pi v_{s}^3}{\alpha^4} p(\eta_{f}) e^{-\alpha 
(t_{f}-t)} & t< t_{f} \\
  0 & t \geq t_{f}
  \end{array} \right.
\end{equation}
where terms $ \exp \left [ -\alpha(t_f-t_c) \right ] $ have been neglected
compared to terms 
$ \exp \left [ -\alpha(t_f-t) \right ] $, because of the steepness of $ p(t) $.
Demanding the whole Universe to have been reheated at $ t_f $ gives an equation
for the maximum amount of supercooling achieved during the transition 
\begin{equation}
\label{etamax}
  \frac{8 \pi v_{s}^3}{\alpha^4} p(\eta_{f}) = 1.
\end{equation}
It has an approximate solution given by
\begin{eqnarray}
\label{etaapp}
 \eta_f & \simeq & \frac{0.4}{\sqrt{2}}\frac{\sigma^{3/2}}{LT_c^{1/2}} 
\sqrt{1-3b\eta_f+\left ( 1-2b\eta_f \right )^{3/2} } \nonumber \\ & 
\simeq & 0.48 \frac{\sigma^{3/2}}{L T_c^{1/2}} \sqrt{1-3b\eta_f}.
\end{eqnarray}
A finite $ \gamma $ approximately halves the critical amount of supercooling,
since it turns out that $ b\eta_f \simeq \frac{1}{4} $. The behavior of
$\eta_f$ is shown in Fig.\ \ref{fig3}. Notice, that 
$ \eta_f $ scales with $ 1/L $.

The approximate solution in Eq.\ (\ref{etaapp}) ignored the details of
the pre-exponential factor, $C$. To see that this is justified,
neglect for a moment the curvature term. It is then easy to show that
\begin{eqnarray}
 \eta_f & = & 4.1\frac{\sigma^{3/2}}{LT_c^{1/2}} \\ \nonumber & \times & 
\left [ 195+\ln C -4\ln \left ( \frac{32\pi \sigma^3 T_c}{3L^2\eta_f^3}
+T_c^2\frac{C'}{C} \right ) \right ]^{-1/2},
\end{eqnarray}
with all dimensional quantities in MeV-units. For the 
$ C(\eta ) $ in Eq.\ (\ref{ct})  $ C'/C=-1/\eta_f $. Because 
\begin{equation}
  \frac{32\pi \sigma^3 T_c}{3L^2\eta_f^3} \gg \frac{T_c^2}{\eta_f},
\end{equation}
when $ \eta_f $ is approximated by Eq.\ (\ref{etaapp}), 
we can safely ignore the $ C'/C $ term. 
Thus a pre-exponential factor of the form Eq.\ (\ref{ct}) 
does not significantly influence
the maximum amount of supercooling neither directly, not even changes in $ C $ 
of several orders of magnitude, nor through its derivative. This result is 
independent of bag model parameters. The same qualitatively picture holds when 
the curvature term is included, if we do nothing but 
substitute Eq.\ (\ref{rc}) for $ R_c $. Notice however, 
that this may not be valid since the
pre-exponential factor is derived without taking a curvature term into account.

The mean nucleation distance is found by substituting Eqs.\ (\ref{faft}) and 
(\ref{paft}) into Eq.\ (\ref{nnuc}) and integrating,
\begin{equation}
\label{dnucex}
 d_{nuc}=(16\pi )^{1/3} \frac{v_s}{\alpha},
\end{equation}
where terms like $ \exp \left [ -\alpha(t_f-t_c) \right ] $ have been neglected.
It should be emphasized that this is also a general result independent of bag
model parameters. Only $ \alpha $ is model dependent.
With $ \alpha $ from Eq.\ (\ref{alfa}), neglecting the $ C'/C $ term, 
$ d_{nuc} $ becomes
\begin{eqnarray}
 d_{nuc} & \simeq & 1.3 \times 10^9    
 \left [ \frac{L^2}{\sigma^3 T_c} {\hbox {MeV}}^2 \right ] \eta_f^3 
  \\ & \times & \left [ \frac{1}{4} \left (2-3b\eta_f +(2-b\eta_f)
\sqrt{1-2b\eta_f}\,\right ) \right ]^{-1} 
\,{\hbox {cm}}. \nonumber
\end{eqnarray}
Eq.\ (\ref{dsurf}) is found by inserting the first approximation for 
$ \eta_f $ from Eq.\ (\ref{etaapp}) with $ \gamma =0 $. 
For $\gamma\neq 0$ one finds in the same manner from the second
approximation in Eq.\ (\ref{etaapp})
\begin{equation}
 d_{nuc} \approx 1.8 \left ( \frac{145\hbox{MeV}}{B^{1/4}} \right )^2 
\left ( \frac{4B}{L} \right )
\left ( \frac{\gamma }{0.24B^{1/2}} \right )^3 \hbox{cm},
\end{equation}
but $ \gamma $ is very insensitive to varying $ s $ quark mass because of the
large contributions from the massless species. Thus the last term is always 
close to unity and we recover Eq.\ (\ref{dcurv}).
Note, the mean nucleation distance is also inversely proportional to the latent 
heat.

In Figure \ref{fig4} we have plotted the mean 
nucleation distance as a function of the 
strange quark mass for a small and a large bag constant with and without the 
curvature term. Curves for intermediate bag constants lie between those two 
curves. This has been done by solving Eq.\ (\ref{etamax}) numerically and 
inserting in Eq.\ (\ref{dnucex}) for $ d_{nuc} $. For comparison the two
approximative solutions, Eqs.\ (\ref{dsurf}) and (\ref{dcurv}) 
have been plotted, too.
If we look at Eq.\ (\ref{dnucex}) we see that for increasing $ \alpha $, i.e.\ 
decreasing amount of supercooling, then the mean nucleation distance decreases.
This is because a larger $ \alpha $ means a more sudden nucleation and more
bubbles are needed in order to reheat the whole Universe at $ \eta_f $. 
Inclusion of a curvature term lowers the energy barrier and thereby the 
maximum amount of supercooling, resulting in roughly a factor of 7 reduction 
of $ d_{nuc} $.

\subsection{Heterogeneous Nucleation}
In a first order phase transition the presence of impurities lowers the energy
barrier and thereby reduces the maximum amount of supercooling achieved during
the transition. We assume that a number density, $ n $, of impurities were 
present in the Universe prior to the quark-hadron transition. Candidates could 
be primordial black holes, magnetic monopoles, cosmic strings, or relic 
fluctuations from the 
electroweak transition. Especially the latter candidate is interesting because 
the horizon distance at the time of the electroweak transition $ \sim 0.3 $ cm 
($ t \sim 10^{-11} $s, $ T \sim 100 $GeV) has expanded three orders of 
magnitude to $ \sim 3 $ meters at the time of the quark-hadron transition.
Further we assume that nucleation around all impurities takes place at a time 
$t_i$, where $t_c \leq t_i < t_f$, and that the bubble expansion is similar to 
the homogeneous case. 

The combined nucleation rate can then be written as 
a sum of the nucleation rate around impurities and the homogeneous nucleation 
rate,
\begin{equation}
 p_{com}(t)=n \delta (t-t_i) + p(\eta_f) e^{-\alpha (t_f-t)},
\end{equation}
where $ \delta $ is Dirac's delta function.
The unaffected fraction of the Universe can still 
be evaluated in the same approximation leading to Eq.\ (\ref{fafta}) although 
the approximation becomes slightly poorer.
\begin{equation}
  f(t) \! \simeq \! \left\{ \! \begin{array}{ll}
  1- \frac {4 \pi v_{s}^3}{3} \left (n(t-t_i)^3 +\! 
\frac{6}{\alpha^4}p(\eta_{f}) e^{-\alpha (t_{f}-t)} \right ) & t< t_{f} \\
  0 & t \geq t_{f}
  \end{array} \right.
\end{equation}
and the equation for $ \eta_f $ now reads
\begin{equation}
\label{etafinh}
  n(t_f-t_i)^3+\frac{6}{\alpha^4}p(\eta_f)=\frac{3}{4\pi v_s^3}.
\end{equation}
The number density of nucleation sites is found from Eq.\ (\ref{nnuc}),
\begin{eqnarray}
\label{nnucinh}
 n_{nuc} & \equiv & d_{nuc}^{-3} \simeq n +\frac{p(\eta_f)}{\alpha} - 
\frac{4\pi v_s^3p(\eta_f)^2}{\alpha^5}  \nonumber \\ & - & 
\frac{4\pi v_s^3}{3}\frac{p(\eta_f)}{\alpha^4}n
 \left [ z^3-3z^2+6z-6 \right ]
\end{eqnarray}
where
$ z=\alpha(t_f-t_i) $. Terms like $ \exp(-\alpha(t_f-t_c)) $ have been 
neglected.
We can write the time difference $ t_f-t_i $ as
\begin{equation}
 t_f-t_i=(1-\xi)(t_f-t_c),
\end{equation}
where $ 0 \leq \xi < 1 $. 

Eq. (\ref{nnucinh}) can be simplified by introducing Eq. (\ref{etafinh}) and 
keeping only the term $ z^3 $ in the parenthesis. Since $ \alpha(t_f-t_c) \sim 
600 $ and we demand $ z^3>10(3z^2-6z+6) $ then we must have $ \xi \leq 0.95 $. 
This is no serious limitation, as we will see. And we hereby decouple the 
heterogeneous and homogeneous contributions to the number density of nucleation 
sites. Eq.\ (\ref{nnucinh}) now reads
\begin{equation}
 d_{nuc}^{-3} \simeq n + 4\pi v_s^3 \frac{p(\eta_f)^2}{\alpha^5}
\simeq n + \frac{\alpha^3}{16\pi v_s^3} .
\end{equation}
In the limits of vanishing and huge $ n $, we recover the expected results, 
$ d_{nuc}=d_{nuc,hom} $ (Eq.\ (\ref{dnucex})) and $ d_{nuc}=n^{-1/3} $, 
respectively.

Because of the steepness of the homogeneous nucleation rate all nucleations 
take place within a 3\% interval in $ \eta $, just below $ \eta_f $. This 
means that either homogeneous {\it or} heterogeneous nucleation determines the 
mean nucleation distance. Only in a {\it very} narrow range are there 
significant contributions from both. 

In Figure \ref{fig5} we have plotted $ d_{nuc} $ 
versus $ n $ for fixed bag constant and different $ s $ quark masses, in the 
most optimistic case where $ \xi \ll 1 $. Notice the step function behavior 
and the broad range, more than two orders of magnitude, of impurity number 
densities, resulting in mean nucleation distances above one meter. 
The largest mean nucleation distance for given parameters is found when 
$ \eta_f \simeq\eta_{f,hom} $. This is a scenario where 
the shock waves from the nucleated hadron bubbles around the impurities 
have reheated the whole Universe 
just before homogeneous nucleation becomes important. The corresponding $ n $ 
can be found from Eq.\ (\ref{etafinh}) by neglecting the homogeneous term,  
\begin{equation}
\label{dnucopt}
 d_{nuc,max}=n^{-1/3} \simeq 6(1-\xi) \left ( \frac{\eta_f}{3\times 10^{-4}} 
\right ) \left ( \frac{100\hbox{MeV}}{T_c} \right )^2 \hbox{m}.
\end{equation}
The corresponding maximal value of $ l $, comoving at $ T=100 $MeV, is found 
by combining Equations (\ref{dnucopt}) and (\ref{l}) into
\begin{equation}
 l_{max} \simeq 5(1-\xi) \left ( \frac{\eta_f}{3\times 10^{-4}} \right )
 \left ( \frac{100\hbox{MeV}}{T_c} \right ) \hbox{m}.
\end{equation}

If $ \xi <1/2 $ and with a typical value of $\eta_f\simeq 3\times
10^{-4}$ (for $L\simeq 4B$; otherwise $\eta_f\propto 1/L$)
one can have significant nucleosynthesis
effects ($l$ of some meters) from impurities for a broad range of $ n $.
For $\xi\rightarrow 1$ the approximations used in
the derivations above break down, but in this regime the presence of
impurities do not lead to large nucleation distances anyway.

The surface tension rises steeply with increasing quark mass for small masses. 
For realistic $ u $ and $ d $ quark masses \cite{particle} $ m_u=5 $MeV and 
$ m_d=10 $MeV the surface tension increases with about 30\% when the $ s $ 
quark contribution is near its maximum. And $ d_{nuc,max} $ increases almost 
70\% to about 11 meters. If we artificially double the surface tension to mimic 
eventual hadron contributions, we gain more than a factor of 3 on the maximum 
mean nucleation distance.

\section{CONCLUSION}
We have calculated the mean nucleation distance in a 
homogeneous nucleation scenario where a 
curvature term contributes to the surface effects.
The framework of the multiple reflection expansion of the MIT bag 
model has been used to calculate the surface tension and curvature term 
self-consistently. Inclusion of a curvature term reduces $ d_{nuc} $ by
about a factor of 7 to less than 2cm.

Furthermore, we have shown that the 
presence of impurities, for a broad range of densities, can result in mean 
nucleation distances above one meter. Such large distances are needed if 
baryon diffusion should not have smeared out generated baryon inhomogeneities 
before the epoch of nucleosynthesis. It must be emphasized that this conclusion 
is qualitatively independent of the model used to describe surface effects. 
We have only assumed that impurities are present, that they lower the energy 
barrier and that the reheating mechanism is the same regardless of how the 
hadron bubbles nucleated. Thus, similar effects could be important in
other first order transitions as well (see e.g. Ref.\ \cite{hoso}). 

Unfortunately there are no reliable theoretical expectations for the
nature or density of impurities, but we note again the coincidence, that the
horizon distance at the electroweak phase transition has expanded to the
interesting range of a few meters at the quark-hadron transition.
 
Eq.\ (\ref{etaapp}) for $ \eta_f $ in the homogeneous nucleation scenario 
tells us how uncertainties in the different parameters influence $ d_{nuc} $, 
which depends linearly on $ \eta_f $. We believe the curvature coefficient is 
well-known, because the major contributions come from gluons and the two 
massless or very light quarks. A significant surface tension contribution from 
the pions would raise $ \eta_f $ and thus result in a larger $ d_{nuc,max} $ 
and an even broader interval of impurity number densities, where the mean 
nucleation distances is above one meter. 

We have throughout used the self-consistently calculated latent heat of $ 4B $,
 even though some recent lattice calculations 
\cite{latent} suggest that 
the latent heat is about ten times smaller. A 
temperature dependent bag constant that increases with increasing temperature 
may reproduce such a result, but no theoretical predictions of such a 
temperature dependence exist. However,  
$ \eta_f $, $d_{nuc}$, and $l$ are all inversely proportional to the latent 
heat. Thus, a tenfold decrease in $ L $ results in a tenfold increase in
$ \eta_f $, $d_{nuc}$, and $l$.

In Ref.\ \cite{alc2} the spectrum of nucleation site separations is calculated. 
A duality between the nucleation sites and large baryon concentration sites is 
assumed and the effect on nucleosynthesis calculated. If impurities play a 
major role, the nucleation site spectrum would look different.
For randomly distributed impurities the spectrum would be broad. A sort of 
regularity in the impurity distribution
would result in a very narrow distribution around the mean 
nucleation distance. Such an approximation is often used in 
actual inhomogeneous nucleosynthesis calculations.

A serious problem with the bag model in the form used in this paper is the 
local minimum in the free energy at about 10 fm, even above the critical 
temperature\cite{mardor}. One might think about removing it by introducing 
further corrections in the density of states related to zero point
energy, color singlet restrictions, etcetera, but no recipe exists at
the moment. We note however, that such effects will be most important
for small radii, and since the maximum of the free energy barrier
typically lies beyond a radius of 50 fm in the results presented above,
the conclusions of the present investigation are not likely to be
affected much.

\acknowledgments
This work was supported in part by the Theoretical Astrophysics Center
under the Danish National Research Foundation, and by the European Human
Capital and Mobility Program.

\begin{figure}
\caption{The latent heat as a function of the $s$ quark mass for 
$B^{1/4}=145$MeV. The uppermost curve includes only pions, whereas the other 
curves from above include mesons up to masses of 547MeV, 892MeV and 1020MeV, 
respectively.}
\label{fig1}
\end{figure}

\begin{figure}
\caption{The free energy as function of bubble radius for different 
temperatures in MeV at $B^{1/4}=145$MeV and $m_s=150$MeV, with curvature. The 
critical temperature gives the solid curve. The dashed curve is for the maximum 
supercooling, $\eta_f$, where almost all hadron bubbles are nucleated with a 
radius, $R_+\simeq 55$fm. The dotted curve is still for $\eta_f$ but with 
$\gamma =0$. The dash-dotted curve corresponds to an even larger supercooling 
where the local minimum at $R_-\simeq 10$fm can no longer be ignored.} 
\label{fig2}
\end{figure}

\begin{figure}
\caption{The maximum supercooling, $\eta_f$, versus $m_s$. Solid curves
are for $B^{1/4}=145$MeV, the upper for $\gamma=0$. Dashed curves are
for $B^{1/4}=245$MeV, the upper for $\gamma=0$. The dotted curves
correspond to the approximation in Eq.\ (\protect\ref{etaapp}).}
\label{fig3}
\end{figure}

\begin{figure}
\caption{The mean nucleation distance as a function of the $s$ quark mass. The 
solid curves are for $B^{1/4}=145$MeV, the uppermost for $\gamma=0$. The dotted 
curves for $B^{1/4}=245$MeV. Dashed curves are the corresponding approximative 
expressions in Eqs.\ (\protect\ref{dsurf}) and (\protect\ref{dcurv}).}
\label{fig4}
\end{figure}

\begin{figure}
\caption{The mean nucleation distance as a function of the impurity number 
density 
for different $s$ quark masses at $B^{1/4}=145$MeV. From left to right the 
masses are 100MeV, 150MeV and 200MeV, respectively.}
\label{fig5}
\end{figure}

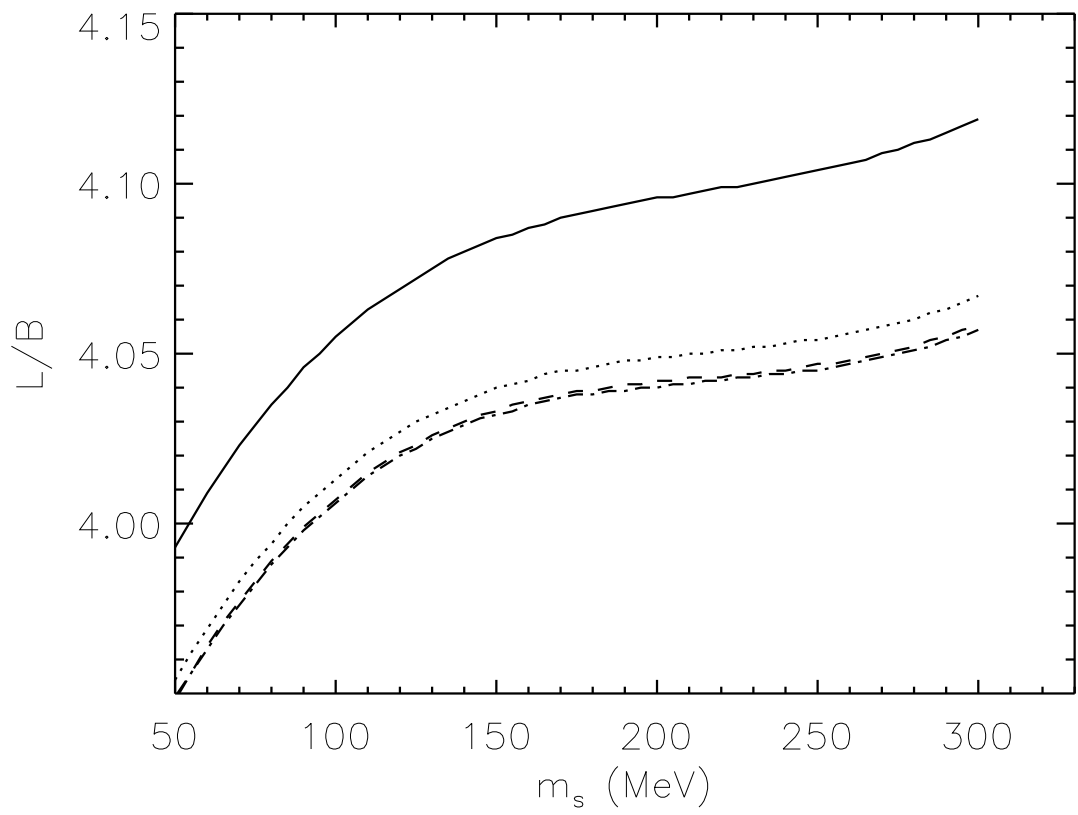
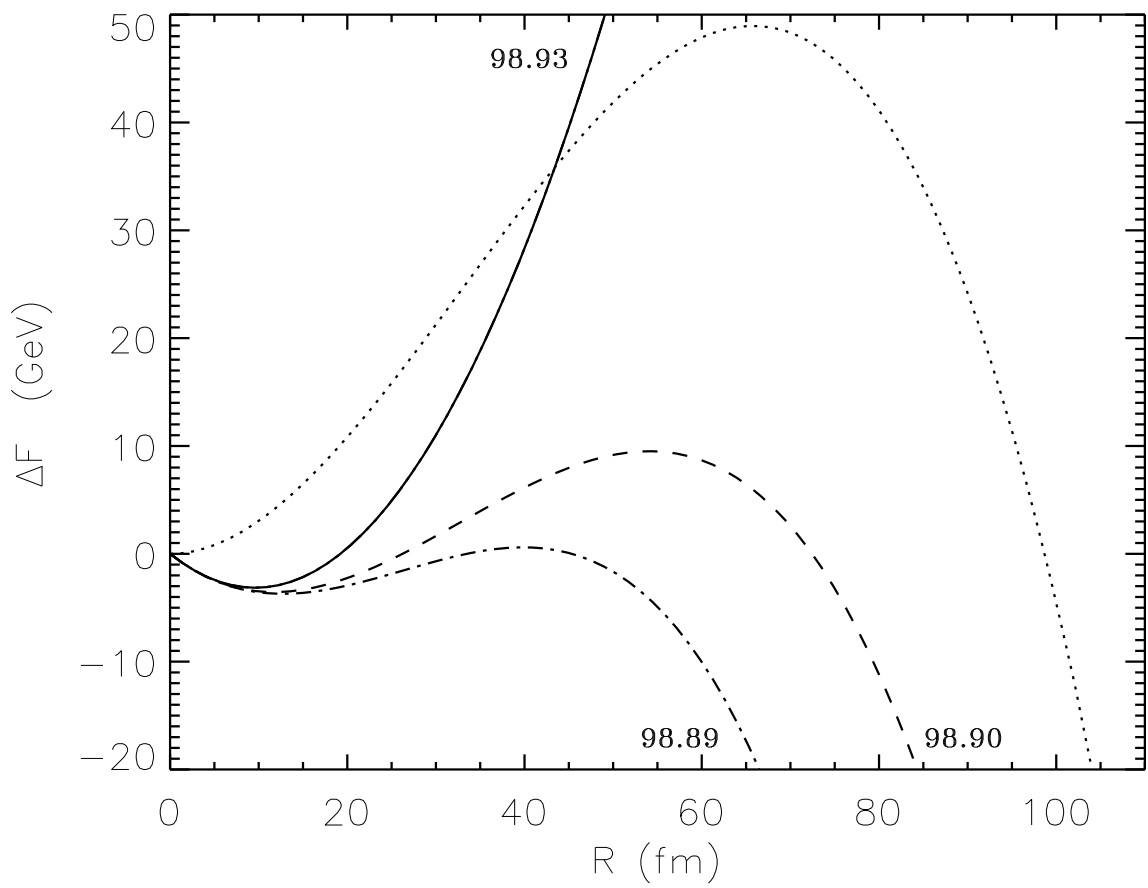
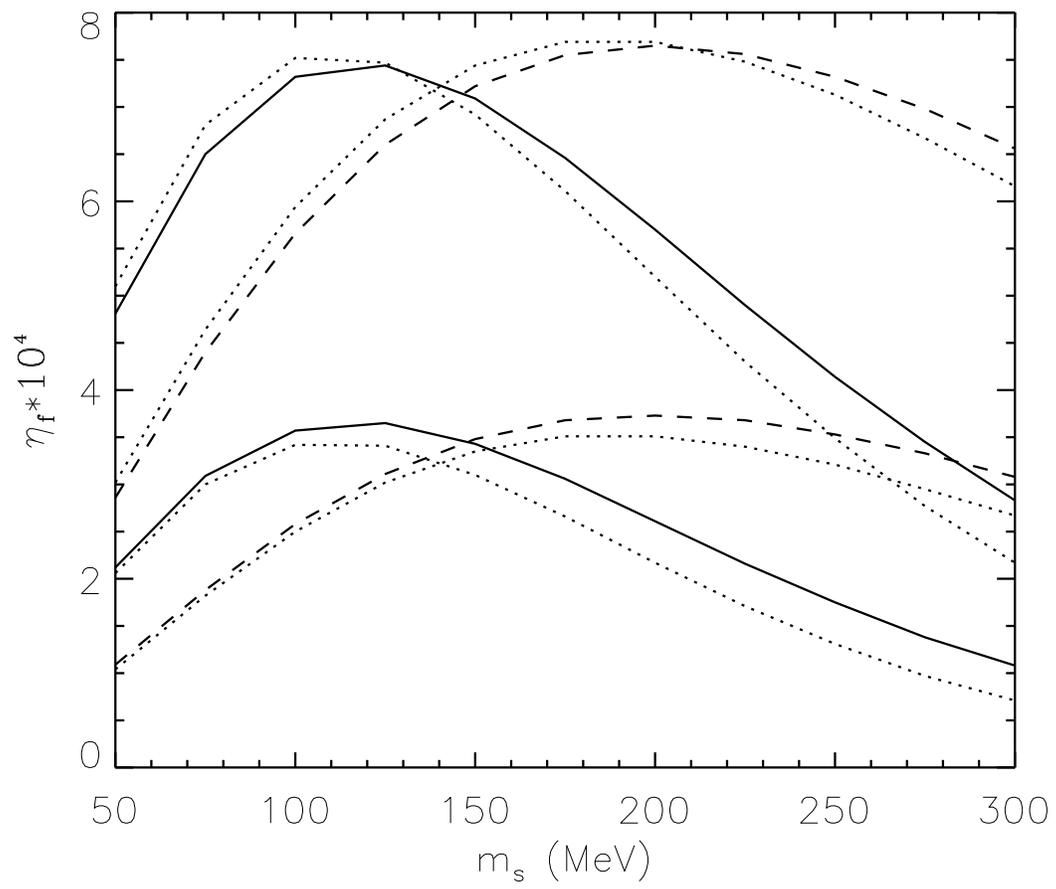
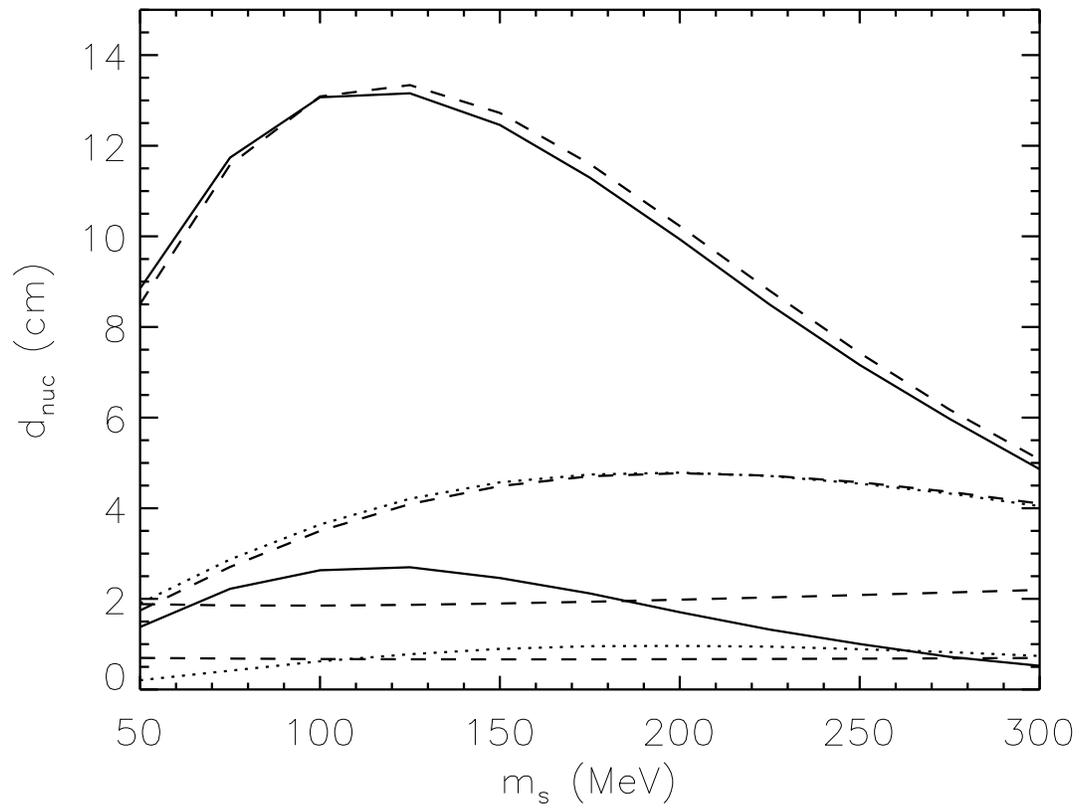
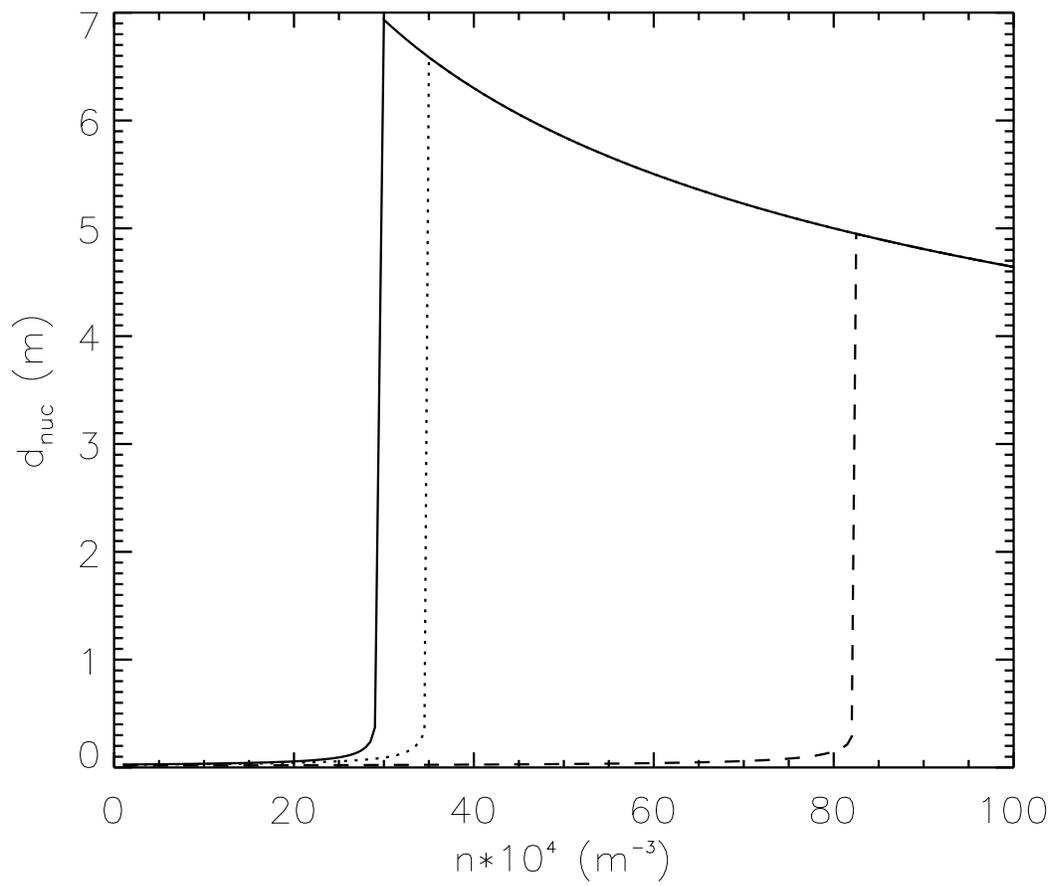
\end{document}